\documentclass{llncs}

\usepackage[normalem]{ulem}

\usepackage{amsmath}
\usepackage{amssymb}
\usepackage{graphicx}

\usepackage{url}
\usepackage{yfonts}
\usepackage{color}

\newcommand{\N}{\mathbb{N}}
\newcommand{\C}{\mathbb{C}}

\newcommand{\x}{\mathbf{x}}
\newcommand{\y}{\mathbf{y}}

\newcommand{\bra}[1]{\left< #1 \right|}
\newcommand{\ket}[1]{\left| #1 \right>}

\newcommand{\iprod}[2]{\langle #1 | #2 \rangle}

\newcommand{\oprod}[2]{| #1 \rangle\langle #2 |}
\newcommand{\rb}{\raisebox{0.5ex}}

\usepackage{hyperref}

\begin{document}
	
\title{On the Unpredictability of Individual Quantum Measurement Outcomes}
	
\author{Alastair A. Abbott\inst{1,2} \and Cristian S. Calude\inst{1} \and Karl Svozil\inst{3}}

\institute{Department of Computer Science, University of Auckland,
Private Bag 92019, Auckland, New Zealand\\ \email{a.abbott,c.calude@auckland.ac.nz}
\and Centre Cavaill\`es, CIRPHLES, \'Ecole Normale Sup\'erieure, 29 rue d'Ulm, 75005 Paris, France
\and Institute for Theoretical Physics,
Vienna  University of Technology,
Wiedner Hauptstrasse 8-10/136,
1040 Vienna,  Austria\\ \email{svozil@tuwien.ac.at}}


%
%
%
%
%
%
%


\maketitle

\begin{abstract}
We develop a general, non-probabilistic model of prediction which is suitable for assessing the (un)predictability of individual physical events.
We use this model to provide, for the first time, a rigorous proof of the unpredictability of a class of individual quantum measurement outcomes, a well-known quantum attribute postulated or claimed for a long time.

We prove that quantum indeterminism---formally modelled as value indefiniteness---is incompatible with the supposition of predictability: \emph{ measurements of value indefinite observables are unpredictable}.
The proof makes essential use of a strengthened form of the Kochen-Specker theorem proven previously to identify value indefinite observables.
This form of  quantum unpredictability, like the Kochen-Specker theorem, relies on three assumptions: 
compatibility with quantum mechanical predictions, non-contextuality, and the value definiteness of observables corresponding to the preparation basis of a quantum state.

We explore the relation between unpredictability and incomputability and show that the unpredictability of individual measurements of a value indefinite quantum observable complements, and is independent of, the global strong incomputability of any sequence of outcomes of this particular quantum experiment.

Finally, we discuss a real model of hypercomputation whose computational power has yet to be determined, as well as further open problems.
\end{abstract}


\section{Introduction}

The outcomes of measurements on a quantum systems are often regarded to be fundamentally unpredictable~\cite{zeil-05_nature_ofQuantum}.
However, such claims are based on intuition and experimental evidence, rather than precise mathematical reasoning.
In order to investigate this view more precisely, both the notion of unpredictability and the status of quantum measurements relative to such a notion need to be carefully studied.

Unpredictability is difficult to formalise not just in the setting of quantum mechanics, but that of classical mechanics too.
Various physical processes from classical chaotic systems to quantum measurement outcomes are often considered unpredictable, and various definitions, both domain specific \cite{Werndl:2009nx} or more general \cite{Eagle:2005ys}, and of varying formality, have been proposed.
For precise claims to be made, the appropriate definitions need to be scrutinised and the results proven relative to specific definitions.

Quantum indeterminism has been progressively formalised via the notion of value indefiniteness in the development of the theorems of Bell~\cite{bell-66} and, particularly,  Kochen and Specker~\cite{kochen1}.
These theorems, which have also been experimentally tested via the violation of various inequalities~\cite{wjswz-98}, express the impossibility of certain classes of deterministic theories.
The conclusion of value indefiniteness from these no-go theorems rests on various assumptions, amounting to the refusal to accept non-classical alternatives such as non-locality and contextual determinism.
And if value indefiniteness is, as often stated, related to unpredictability, any claims of unpredictability need to be similarly evaluated with respect to, and seen to be contingent on such assumptions.

In this paper we address these issues in turn.
We first discuss various existing notions of predictability and their applicability to physical events.
We propose a new formal model of prediction which is non-probabilistic and, we argue, captures the notion that an arbitrary single physical event (be it classical, quantum, or otherwise) or sequence thereof is `in principle' predictable.
We review the formalism of value indefiniteness and the assumptions of the Kochen-Specker theorems (classical and stronger forms), and show that the outcomes of measurements of value indefinite properties are indeed unpredictable with respect to our model.
Thus, in this framework unpredictability  rests on the same assumptions as quantum value indefiniteness.
Finally, we discuss the relationship between quantum randomness and unpredictability, and show that unpredictability does not, in general, imply the incomputability of sequences generated by repeating the experiment \emph{ad infinitum}.
Thus, the strong incomputability of sequences of quantum measurement outcomes appears to rest independently on the assumption of value indefiniteness.

\section{Models of prediction}

To predict---in  Latin pr{\ae}dicere, ``to say before"---means \emph{to forecast what will happen under specific conditions before the phenomenon happens.}
Various definitions of predictability  proposed by different authors will be discussed regarding their suitability for capturing the notion of predictability of individual physical events or sequences thereof in the most general sense.
While some papers, particularly in physics and cryptographic fields, seem to adopt the view that probabilities mean unpredictability \cite{Acin:2013qa,zeil-05_nature_ofQuantum}, this is insufficient to describe unpredictable physical processes.
Probabilities are a formal description given by a particular theory, but do not entail that a physical process is fundamentally, that is, ontologically, indeterministic nor unpredictable, and can (often very reasonably) represent simply an epistemic lack of knowledge or underdetermination of the theory.
Instead, a more robust way to formulate prediction seems to be in terms of a `predicting agent' of some form.
This is indeed the approach taken by some definitions, and that we also will follow.

In the theory of dynamical systems, unpredictability has long been linked to chaos and has often been identified as the inability to calculate with any reasonable precision the state of a system given a particular observable initial condition~\cite{Werndl:2009nx}.
The observability is critical, since although a system may presumably have a well-defined initial state (a point in phase-space), any observation yields an interval of positive measure (a region of phase space).
This certainly seems the correct path to follow in formalising predictability, but more generality and formalism is needed to provide a definition for arbitrary physical processes.

Popper, in arguing that unpredictability \emph{is} indeterminism, defines prediction in terms of ``physical predicting machines''~\cite{popper-50i}.
He considers these as real machines that can take measurements of the world around them, compute via physical means, and output (via some display or tape, for example) predictions of the future state of the system.
He then studies experiments which must be predicted with a certain accuracy and considers these to be predictable if it is \emph{physically} possible to construct a predictor for them.

Wolpert \cite{Wolpert:2008aa} formalised this notion much further in developing a general abstract model of physical inference.
Like Popper, Wolpert was interested in investigating the limits of inference, including prediction, arising from the simple fact that any inference device must itself be a physical device,  hence an object whose behaviour we can try to predict.
While Wolpert's aim was not so focused on the predictability arising from the nature of specific physical theories, he identified and formalised the need for an experimenter to develop prediction techniques and initialise them by interacting with the environment via measurements.

A more modern and technical definition of unpredictability was given by Eagle~\cite{Eagle:2005ys} in defining randomness as maximal unpredictability.
While we will return to the issue of randomness later, Eagle's definition of unpredictability deserves further attention.
He defined prediction relative to a particular theory and for a particular predicting agent, an approach thus with some similarity to that of Wolpert.
Specifically, a prediction function is defined as a function mapping the state of the system described by the theory and specified epistemically (and thus finitely) by the agent to a probability distribution of states at some time.
This definition formalises more clearly prediction as the output of a function operating on information extracted about the physical system by an agent.

Popper's and Wolpert's notions of predictability perhaps lack generality by requiring the predictor
to be embedded, that is, physically present, in its environment~\cite{toffoli:79}, and are not so suited to investigating the predictability of particular physical processes, but rather of the physical world as a whole.
Similarly, Eagle's definition renders predictability relative to a particular physical theory.

In  order to relate the intrinsic indeterminism of a system to unpredictability, it would be more appropriate to have a definition of events as unpredictable \emph{in principle}.
Thus, the predictor's ignorance of a better theory might change their associated epistemic ability to know if an event is predictable or not, but would not change the fact that an event may or may not be, in principle, predictable.

Last but not least, it is important to restrict the class of prediction functions by imposing some effectivity (i.e. computability) constraints. 
Indeed, we suggest  that ``to predict''  is to say in advance in some effective/constructive/computable way what physical event or outcome will happen. 
Thus, motivated by the Church-Turing Thesis, we choose here Turing computability.
Any predicting agent operating with incomputable means---incomputable/infinite inputs or procedures that can go beyond the
the power of algorithms (for example, by executing infinitely many operations in a finite amount of time)---seems to be physically highly speculative if not impossible.
Technically, ``controlled incomputability'' could be easily incorporated in the model, if necessary.

Taking these points into account, we propose a definition---similar in some aspects to Wolerpt's and Eagle's definitions---based on the ability of some computably operating agent to correctly predict using finite information extracted from the system of the specified experiment. 
For simplicity we will consider experiments with binary observable values (0 or 1), but the extension to finitely or countable many (i.e. finitely specified) output values is straightforward.
Further, unlike Eagle~\cite{Eagle:2005ys}, we consider only prediction with certainty, rather than with probability.
While it is not difficult nor  unreasonable to extend our definition to the more general scenario, this is not needed for our application to quantum measurements;
moreover,  in doing so we avoid any potential pitfalls related to probability 1 or 0 events \cite{Zaman:1987gd}.

Our main aim is to define the (correct) prediction of individual events \cite{Eagle:2005ys}, which can be easily extended to an infinite sequence of events.
An individual event can be correctly predicted simply by chance, and a robust definition of predictability clearly has to  avoid this possibility.
Popper succinctly summarises this predicament in Ref.~\cite[117--118]{popper-50i}:
\begin{quote}
``\emph{If we assert of an observable event that it is unpredictable we do not mean, of course, that it is logically or physically impossible for anybody to give a correct description of the event in question before it has occurred;
for it is clearly not impossible that somebody may hit upon such a description accidentally.
What is asserted is that certain rational methods of prediction break down in certain cases---the methods of prediction which are practised in physical science.}''
\end{quote}

One possibility is then to demand a proof that the prediction is correct,  thus formalising the ``rational methods of prediction'' that Popper refers to.
However, this is notoriously difficult and must be made relative to the physical theory considered, which generally is not well axiomatised and can change over time.
Instead we demand that such predictions be \emph{repeatable}, and not merely one-off events.
This point of view is consistent with Popper's own framework of empirical falsification~\cite{popper,popper-en}: an empirical theory (in our case, the prediction) can never be proven correct, but it can be falsified through decisive experiments pointing to incorrect predictions.
Specifically, we require that the \emph{predictions remain correct in any arbitrarily long (but finite) set of repetitions of the experiment.}

\section{A model for prediction of individual physical events}

In order to formalise our non-probabilistic model of prediction we consider a hypothetical experiment $E$ specified effectively by an experimenter.
We formalise the notion of a predictor as an effective (i.e.\ computational) method of uniformly producing the outcome of an experiment using finite information extracted (again, uniformly) from the experimental conditions along with the specification of the experiment, but \emph{independent} of the results of the experiments. 
An experiment will be predictable if any potential sequence of repetitions (of unbounded, but finite,  length) of it can always be predicted correctly by such a predictor.

In detail, we consider a finitely specified physical experiment $E$ producing a single bit $x\in\{0,1\}$ (which, as we previously noted, can readily be generalised).
Such an experiment could, for example, be the measurement of a photon's polarisation after it has passed through a 50-50 polarising beam splitter, or simply the toss of a physical coin with initial conditions and experimental parameters specified finitely.
Further, with a particular instantiation or trial of $E$ we associate the parameter $\lambda$  which fully describes the trial.
While $\lambda$ is not in its entirety an obtainable quantity, it contains any information that may be pertinent to prediction and any predictor can have practical access to a finite amount of this information.
In particular this information may be   directly associated with the particular trial of $E$ (e.g. initial conditions or hidden variables) and/or relevant external factors (e.g. the time, results of previous trials of $E$). We can view $\lambda$ as a resource that one can extract finite information from in order   to predict the outcome of the experiment $E$.
Any such external factors should, however, be local in the sense of special relativity, as (even if we admit quantum non-locality) any other information cannot be utilised for the purpose of prediction~\cite{laloe-2012}.
We formalise this in the following.

An \emph{extractor} is a physical device selecting a finite amount of information included in $\lambda$ without altering the experiment $E$.
It can be used by a predicting agent to examine the experiment and make predictions when the experiment is performed with parameter $\lambda$. 
Mathematically, an extractor is represented by a (deterministic) function $\lambda \mapsto \xi(\lambda)\in\{0,1\}^*$ where $\xi(\lambda)$ is a finite string of bits. 
For example, $\xi(\lambda)$ may be an encoding of the result of the previous instantiation of $E$, or the time of day the experiment is performed. 
As usual, the formal model is significantly weaker: here, an extractor is a deterministic function which can be physically implemented without affecting the experimental run of $E$.

A predictor for $E$ is an algorithm  (computable function) $P_E$ which \emph{halts} on every input and \emph{outputs} either $0$, $1$ (cases in which  $P_E$ has made a prediction), or ``prediction withheld''.
We interpret the last form of output as a refrain from making a prediction.
The predictor $P_E$ can utilise as input the information $\xi(\lambda)$ selected by an extractor
encoding  relevant information for a particular instantiation of $E$, but must not disturb or interact with $E$ in any way;
that is, it must be \emph{passive}.

As we noted earlier, a certain predictor may give the correct output for a trial of $E$ simply by chance.
This may be due not only to a lucky choice of predictor, but also to the input being chosen by chance to produce the correct output.
Thus, we rather  consider the performance of a predictor $P_E$  using, as input, information extracted by a particular fixed extractor.
This way we ensure that $P_E$ utilises in ernest information extracted from $\lambda$,
and we avoid the complication of deciding under what input we should consider $P_E$'s correctness.

A predictor $P_E$ provides a \emph{correct prediction} using the extractor $\xi$ for an instantiation of $E$ with parameter $\lambda$ if, when taking as input $\xi(\lambda)$, it outputs 0 or 1 (i.e.\ it does not refrain from making a prediction) and this output is equal to $x$, the result of the experiment.

Let us fix an extractor $\xi$. The predictor $P_E$ is \emph{$k$-correct for $\xi$} if there exists an $n\ge k$ such that when $E$ is repeated $n$ times with associated parameters $\lambda_1 ,\dots, \lambda_n$ producing the outputs $x_1,x_2,\dots ,x_n$, $P_E$ outputs the sequence $P_E(\xi(\lambda_1)), P_E(\xi(\lambda_2)),\dots ,P_E(\xi(\lambda_n))$ with the following two properties:
\begin{enumerate}
\item no prediction in the sequence is incorrect, and
\item in the sequence there are $k$ correct predictions.
\end{enumerate}
The repetition of $E$ must follow an algorithmic procedure for resetting and repeating the experiment;
generally this will consist of a succession of events of the form ``$E$ is prepared, performed, the result (if any)  recorded, $E$ is reset''.

If $P_E$ is $k$-correct for $\xi$ we can bound the probability that $P_E$ is in fact operating by chance and may not continue to give correct predictions, and thus give a measure of our confidence in the predictions of $P_E$.
Specifically, the sequence of $n$ predictions made by $P_E$ can be represented as a string of length $n$ over the alphabet $\{T,F,W\}$, where $T$ represents a correct prediction, $F$ an incorrect prediction, and $W$ a withheld prediction.
Then, for a predictor that is $k$-correct for $\xi$ there exists an $n\ge k$ such that the sequence of predictions contains $k$ $T$'s and $(n-k)\,$ $W$'s.
There are ${n \choose k}$ such possible prediction sequences out of $3^n$ possible strings of length $n$.
Thus, the probability that such a correct sequence would be produced by chance tends to zero when $k$ goes to infinity because
$$\frac{{n\choose k}}{3^n}<\frac{2^n}{3^n}\le \left(\frac{2}{3}\right)^k\rb.$$

Clearly the confidence we have in a $k$-correct predictor increases as $k\to\infty$.
If $P_E$ is $k$-correct for $\xi$ for all $k$, then $P_E$ never makes an incorrect prediction and the number of correct predictions can be made arbitrarily large by repeating $E$ enough times.
In this case, we simply say that \emph{$P_E$ is correct for $\xi$}.
The infinity used in the above definition is \emph{potential} not
actual: its role is to guarantee arbitrarily many correct predictions.

This definition of correctness allows $P_E$ to refrain from predicting when it is unable to.
A predictor $P_E$ which is correct for $\xi$ is, when using the extracted information $\xi(\lambda)$, guaranteed to always be capable of providing more correct predictions for $E$, so it will not output ``prediction withheld'' indefinitely.
Furthermore, although $P_E$ is technically used only a finite, but arbitrarily large, number of times, the definition guarantees that, in the hypothetical scenario where it is executed infinitely many times, $P_E$ will provide  infinitely many correct predictions and not a single incorrect one.

While a predictor's correctness is based on its performance in repeated trials,  we can use the predictor to define the prediction of single bits produced by the experiment $E$.
If $P_E$ is not correct for $\xi$ then we cannot exclude the possibility that any correct prediction $P_E$ makes is simply due to chance.
Hence, we propose the following definition:
\begin{quote}
\emph{the outcome $x$ of a single trial of the experiment $E$ performed with parameter $\lambda$ is {\rm predictable} (with certainty) if there exist an extractor $\xi$ and a predictor $P_E$ which is correct for $\xi$, and $P_E(\xi(\lambda))=x$}.
\end{quote}

Accordingly, $P_{E}$ correctly predicts the outcome $x$, never makes an incorrect  prediction, and can produce arbitrarily many correct predictions.

\section{Computability theoretic notions of unpredictability}
\label{sec:alg}

The notion of unpredictability defined in the previous section has both physical components (in extracting information from the system for prediction via $\xi$) and computability theoretic ones (in predicting via an effective procedure, $P_E$).
Both these components are indispensable for a good model of prediction for physical systems, but it is nonetheless important to discuss their relation to pure computability theoretic notions of prediction, since these place unpredictability in a context where the intuition is stripped to its abstract basics.

The algorithmic notions of bi-immunity (a strong form of incomputability) and Martin-L\"of randomness describe some forms of unpredictability for infinite sequences of bits~\cite{calude:02}.
A sequence is \emph{bi-immune} if it contains no infinite computable subsequence (i.e., both the bits of the subsequence and their positions in the original sequence must be computable).
A sequence is  \emph{Martin-L\"of random} if 
all prefixes 
of the sequence cannot be compressed by more than an additive constant  by a universal prefix-free Turing machine (see~\cite{calude:02,DH} for more details).
Thus, for a bi-immune sequence, we cannot effectively compute the value of any bit in advance and only finitely many bit-values can be correctly ``guessed'', while a Martin-L\"of random sequence contains no ``algorithmic" patterns than can be used to effectively compress it.

However, the notions of predictability presented by Tadaki~\cite{Tadaki14} are perhaps the most relevant for this discussion.
\textit{An infinite sequence of bits $\x=x_1x_2\dots$ is \emph{Tadaki totally predictable} if there exists a Turing machine $F:\{0,1\}^*\to\{0,1,W\}$ that halts on every input, and satisfies the following two conditions: (i) for every $n$, either $F(x_1\dots x_n)=x_{n+1}$ or $F(x_1\dots x_n)=W$; and (ii) the set $\{n\in \N^+ \mid F(x_1\dots x_n)\neq W\}$ is infinite;} $F$ is called a \emph{total predictor} for $\x$.

A similar notion, called \emph{Tadaki predictability}, requires only that $F$ halts on all input $x_1\dots x_n$, and thus may be a partially computable function instead of a computable one.
This emphasises that, as we mentioned earlier, the notion of predictability can be strengthened or weakened by endowing the predictor with varying computational powers.

Tadaki predictability can be related to various other algorithmic notions of randomness.
For example, \emph{no Martin-L\"of random sequence is Tadaki (totally) predictable}~\cite[Theorem~4]{Tadaki14}, while \emph{all non-bi-immune sequences are Tadaki totally predictable}.
This last fact can be readily proven by noting that a non-bi-immune sequence $\x$ must contain a computable subsequence $(k_1,x_{k_1}),(k_2,x_{k_2}),\dots$.
Equivalently, there is an infinite computable set $K\subset \N$ and a computable function $f:K\to \{0,1\}$ such that for all $k\in K$, $f(k)=x_k$.
Hence, for a string  $\sigma\in\{0,1\}^*$ the function $$F(\sigma)=\begin{cases} f(|\sigma|+1), & \text{if $|\sigma|+1\in K$,}\\ W, & \text{otherwise,}\end{cases}$$ is  a Tadaki total predictor for $\x$ ($|\sigma|$ is the length of $\sigma $).

Furthermore, the notion of \emph{Tadaki total unpredictability is strictly stronger than bi-immunity}, since there exist bi-immune, totally predictable sequences.
For example, let $\x=x_1x_2\dots$ be a Martin-L{\"o}f random sequence (and hence bi-immune~\cite{calude:02}).
It is not difficult to show that $\y=y_1y_2\dots = x_1 x_1 x_2 x_2\dots$ created by doubling the bits of $\x$ is bi-immune.
However, $\y$ has a Tadaki total predictor $F$ defined as $$F(\sigma_1 \dots \sigma_n) = \begin{cases} \sigma_n, & \text{if $n$ is odd,}\\ W, & \text{if $n$ is even,}\end{cases}$$ since this correctly predicts the value of every bit at an even position in $\y$.

This notion of predictability can be physically interpreted in the following way.
Consider a black-box $B(\x)$ with a button that, when pressed, gives the next digit of $\x$; by repeating this operation one can slowly learn, in order, the bits of $\x$.
A sequence is Tadaki predictable if there is a uniform way to compute infinitely often $x_{n+1}$ having learnt the initial segment $x_1\dots x_n$, with the proviso that we must know \emph{in advance} when---that is, the times at which---we will be able to do so.

When viewed from the physical point of view described above, there is a clear relation to our notion of predictability.
In particular, we can consider a deterministic experiment $E_{\x}$ that consists of generating a bit from the black-box $B(\x)$, and asking if $E_{\x}$ is predictable for the `prefix' extractor $\xi_p(\lambda_{i})=x_1\dots x_{i-1}$ for the trial of $E_{\x}$ producing $x_{i}$---that is, using just the results of the previous repetitions of $E_{\x}$.
It is not too difficult to see that \emph{$E_{\x}$ is predictable if and only if $\x$ is Tadaki totally predictable}.
Indeed,  equate the function $F$ from Tadaki's definition and the predictor $P_E$, as well as the outputs `$W$' and ``prediction withheld''.

In general, algorithmic information theoretical properties of sequences could be explored using our model of prediction via such an approach.
However, the relation between these notions exists only when one considers particular, abstract, extractors such as $\xi_p$.
The generality of our model originates in the importance it affords to physical properties of systems, \emph{via} extractors, which are essential for prediction in real systems.
Depending on the physical scenario investigated, then, physical devices might allow us to extract information allowing  to predict an experiment, regardless of the algorithmic content of this information, as long as finite information suffices for a single prediction.

\section{Quantum unpredictability}

We now  apply the notion developed above to formally justify  the well-known claim that quantum events are completely unpredictable.

\subsection{The intuition of quantum indeterminism and unpredictability}

Intuitively, it would seem that quantum indeterminism corresponds to the \emph{absence of physical reality};
if no unique element of physical reality corresponding to a particular physical quantity exists, this is reflected by the physical quantity being indeterminate.
That is, for such an observable none of the possible exclusive measurement outcomes are certain to occur and therefore we should conclude that any kind of prediction of the outcome with certainty cannot exist, and the outcome of this individual measurement must thus be unpredictable.  
For example, an agent trying to predict the outcome of a measurement of a projection observable in a basis unbiased with respect to the preparation basis (i.e.\  if there is a ``maximal mismatch'' between preparation and measurement) could do no better than blindly guess the outcome of the measurement.

However, such an argument is  too informal.  To apply our model of unpredictability 
the notion of indeterminism needs to be specified much more rigorously: this implies developing a formalism for quantum indeterminism, as well as a careful discussion of the assumptions which indeterminism is reliant on.

\subsection{A formal basis for quantum indeterminism}
\label{sec:FQI}

The phenomenon of quantum indeterminism cannot be deduced from the Hilbert space formalism of quantum mechanics alone, as this specifies only the probability distribution for a given measurement which in itself need not indicate intrinsic indeterminism.
Indeterminism has had a role at the heart of quantum mechanics since Born postulated that the modulus-squared of the wave function should be interpreted as a probability density that, unlike in classical statistical physics~\cite{Myrvold2011237}, expresses fundamental,  irreducible indeterminism~\cite{born-26-1}.
In Born's own words, ``\emph{I myself am inclined  to give up determinism in the world of atoms.}''
The nature of individual measurement outcomes in quantum mechanics was, for a period, a subject of much debate.
Einstein famously dissented, stating his belief that \cite[p. 204]{born-69} ``\emph{He does not throw dice}.''
Nonetheless, over time the conjecture that measurement outcomes are themselves fundamentally indeterministic became the quantum orthodoxy~\cite{zeil-05_nature_ofQuantum}.

Beyond the blind belief originating with Born, the Kochen-Specker theorem, along with Bell's theorem, are among the primary reasons for the general acceptance of quantum indeterminism.
The belief in quantum indeterminism thus rests largely on the same assumptions as these theorems.
In the development of the Kochen-Specker theorem, quantum indeterminism has been formalised as the notion of value indefiniteness~\cite{2012-incomput-proofsCJ}, which allows us to discuss indeterminism in a more general formal setting rather than restricting ourselves to any particular interpretation.
Here we will review this formalism, as well as a stronger form of the Kochen-Specker theorem and its assumptions which are important for the discussion of unpredictability.

For a given quantum system in a particular state, we say that an observable is \emph{value definite} if the measurement of that observable is pre-determined to take a (potentially hidden) value.
If no such pre-determined value exists, the observable is \emph{value indefinite}.
Formally, this notion can be represented by a \emph{(partial) value assignment function} (see~\cite{2012-incomput-proofsCJ} for the complete formalism).

In addressing the question of when we should conclude that a physical quantity is value definite, Einstein, Podolsky and Rosen (EPR) give \emph{a sufficient criterion of physical reality} in terms of certainty and predictability in \cite[p.~777]{epr}.
Based on this accepted \emph{sufficient} condition for the existence of an element of physical reality, we allow ourselves to be guided by the following  ``EPR 
principle'':\footnote{They continue: ``It seems to us that this criterion, while far from exhausting all possible ways of recognizing a physical reality, at least provides us with one such way, whenever the conditions set down in it occur.''}

\begin{quote}
	\emph{EPR principle}: If, without in any way disturbing a system, we can predict with certainty the value of a physical quantity, then there exists a \emph{definite value} prior to observation corresponding to this physical quantity.
\end{quote}

As we discussed earlier, the notion of prediction the EPR principle refers to needs to be effective; further, we
 note that the constraint that prediction acts ``without in any way disturbing a system'' is perhaps non-trivial~\cite{laloe-2012}, but is equally required by our model of prediction.

The EPR principle justifies the subtle but often overlooked 
\begin{quote}\emph{Eigenstate principle}:
If a quantum system is prepared in a state $\ket{\psi}$, then the projection observable $P_\psi=\oprod{\psi}{\psi}$ is value definite.
\end{quote}
This principle is necessary in order to use the strong Kochen-Specker theorem to single-out value indefinite observables, and is similar to, although weaker, than the eigenstate-eigenvalue link (as only one direction of the implication is asserted)~\cite{Suarez:2004gn}.

 A further requirement called \emph{admissibility} is used to avoid outcomes  impossible to obtain according to quantum predictions.
Formally, admissibility states that an observable in a context---that is, a set of mutually commuting (i.e. compatible) observables---cannot be value indefinite if all but one of the possible measurement outcomes would contradict quantum mechanical identities given the values of other, value definite observables in the same context.
In such a case, the observable must have the definite value of that sole `consistent' measurement outcome.

Here is an example: given a context $\{P_1,\dots,P_n\}$ of commuting projection observables, if $P_1$ were to have the definite value 1, all other observables in this context must have the value 0.
Were this not the case, there would be a possibility to obtain the value 1 for more than one compatible projection observable, a direct contradiction of the quantum prediction that one and only one projector in a context give the value 1 on measurement.
Note that we require this to hold only when any indeterminism (which implies multiple possible outcomes) would allow quantum mechanical predictions to be broken:
were $P_1$ to have the value 0, admissibility would not require anything of the other observables if the rest were value indefinite, as neither a measurement outcome of 0 or 1 for $P_2\dots P_n$ would lead to a contradiction.

The Kochen-Specker theorem \cite{kochen1} shows that no value assignment function can consistently make \emph{all} observables value definite while maintaining the requirement that the values are assigned non-contextually---that is, the value of an observable is the same in each context it is in.
This is a global property: non-contextuality is incompatible with \emph{all} observables being value definite.
However, it is possible to go deeper and localise value indefiniteness to prove that even the existence of two non-compatible value definite observables is in contradiction with admissibility and the requirement that any value definite observables behave non-contextually, without requiring that all observables be value definite.
Thus, any mismatch between preparation and measurement context leads to the measurement of a value indefinite observable: this is stated formally in the following strong version of the Kochen-Specker theorem.

\begin{theorem}[From \cite{2012-incomput-proofsCJ,PhysRevA.89.032109}]
	\label{thm:vi-everywhere}
		Let there be a quantum system prepared in the state
	$\ket{\psi}$ in dimension $n\ge 3$ Hilbert space $\C^n$, and let $\ket{\phi}$ be any state neither orthogonal nor parallel to $\ket{\psi}$, i.e.\ $0<|\iprod{\psi}{\phi}|<1$.
	Then the projection observable $P_\phi=\ket{\phi}\bra{\phi}$ is value indefinite under any non-contextual, admissible value assignment.
\end{theorem}

Hence,  accepting that definite values, \emph{should they exist} for certain observables, behave non-contextually is in fact enough to derive rather than postulate quantum value indefiniteness.

\subsection{Contextual alternatives}
\label{sec:ac}

It is worth keeping in mind that, while indeterminism is often treated as an assumption or aspect of the orthodox viewpoint \cite{born-26-1,zeil-05_nature_ofQuantum}, this usually rests implicitly on the deeper assumptions (mentioned in
Sect.~\ref{sec:FQI})  that the Kochen-Specker theorem relies on.
If these assumptions are violated, deterministic theories could not be excluded, and the status of value indefiniteness and unpredictability would need to be carefully revisited.

If this were the case, perhaps the simplest alternative would be the explicit assumption of (albeit non-local) context dependant predetermined values.
Many attempts to interpret quantum mechanics deterministically, such as Bohmian mechanics~\cite{Bohm52}, can be expressed in this framework.
Since such a theory would no longer be indeterministic, the intuitive argument for unpredictability would break down, and the theory could in fact be totally predictable.
However, predictability is still not an immediate consequence, as such hidden variables could potentially be ``assigned'' by a demon operating beyond the limits of any predicting agent (e.g. incomputably).

Another possibility would be to consider the case that any predetermined outcomes may in fact not be determined by the observable alone, but rather by \emph{``the complete disposition  of the apparatus''} \cite[Sec.~5]{bell-66}.
In this viewpoint, even when the macroscopic measurement apparatuses are still idealised as being perfect, their many degrees of freedom (which may by far exceed Avogadro's  or Loschmidt's constants) contribute to any measurement of the single quantum.
Most of these degrees of freedom might be totally uncontrollable by the experimenter, and may result in an \emph{epistemic unpredictability} which is dominated by the combined complexities of interactions between the single quantum measured and the (macroscopic) measurement device producing the outcome.

In such a measurement, the pure single quantum and the apparatus would become entangled.
In the absence of one-to-one uniqueness between the macroscopic states of the measurement apparatus and the quantum, any measurement would amount to a partial trace resulting in a mixed state of the apparatus, and thus to uncertainty and unpredictability of the readout.
In this case, just as for irreversibility in classical statistical mechanics~\cite{Myrvold2011237}, the unpredictability of single quantum measurements might not be irreducible at all, but  an expression of, and relative to, the limited means available to analyse the situation.

\subsection{Unpredictability of  individual quantum measurements}
\label{sec:physUnpred}

With the notion of value indefiniteness presented, let us now turn our attention to applying our formalism of unpredictability to quantum measurement outcomes of the type discussed in  Sect.~\ref{sec:FQI}.

Throughout this section we will consider an experiment $E$ performed in dimension $n\ge 3$ Hilbert space in which a quantum system is prepared in a state $\ket{\psi}$ and a value indefinite observable $P_\phi$ is measured producing a single bit $x$.
By Theorem~\ref{thm:vi-everywhere} such an observable is guaranteed to exist, and to identify one we need only a mismatch between preparation and observation contexts.
The nature of the physical system in which this state is prepared and the experiment performed is not important, whether it be photons passing through generalised beam splitters~\cite{rzbb}, ions in an atomic trap, or any other quantum system in dimension $n\ge 3$ Hilbert space.

We first show that experiments utilising quantum value indefinite observers cannot have a predictor which is correct for some $\xi$.
More precisely: 
\begin{theorem} \label{unpredict}If $E$ is an experiment measuring  a quantum value indefinite observable, then for every predictor $P_E$ using any extractor $\xi$, $P_E$ is not correct for $\xi$.
\end{theorem}

Let us fix an extractor $\xi$, and assume for the sake of contradiction that there exists a predictor $P_E$ for $E$ which is correct for $\xi$.
Consider the  case when the experiment $E$ is repeatedly initialised, performed and reset an arbitrarily large but finite, number of times in an algorithmic ``ritual'' generating a finite sequence of bits $x_1x_2\dots x_n$. 

Since $P_E$ \emph{never} makes an incorrect prediction, each of its predictions is correct with certainty.
Then, according to the EPR principle we must conclude that each such prediction corresponds to a value definite property of the system measured in $E$.
However, we chose $E$ such that this {\it  is not}  the case: each $x_i$ is the result of the measurement of a value indefinite observable, and thus we obtain a contradiction and conclude no such predictor $P_E$ can exist.

Moreover, since there does not exist a predictor $P_E$ which is correct for some $\xi$, for such a quantum experiment $E$, no single outcome is predictable with certainty.
\begin{theorem} \label{unpredictsingle}
If the experiment $E$ described above is repeated  a) an arbitrarily large, but finite number of times producing the finite sequence $x_1x_2\dots \x_n$, or b) hypothetically, \emph{ad infinitum}, generating the infinite sequence $\x=x_1x_2\dots$, then no single bit $x_i$ can be predicted with certainty.
\end{theorem}

\section{Incomputability, unpredictability, and quantum randomness}

While there is a clear intuitive link between unpredictability and randomness, it is an important point that the unpredictability of quantum measurement outcomes should not be understood to mean that quantum randomness is ``truly random''.
Indeed, the subject of randomness is a delicate one: randomness can come in many flavours~\cite{DH}, from statistical properties to computability theoretic properties of outcome sequences.
For physical systems, the randomness of a process also needs to be differentiated from that of its outcome.

As mentioned earlier, Eagle has argued that a physical process is random if it is ``maximally unpredictable'' \cite{Eagle:2005ys}.
In this light it may be reasonable to consider quantum measurements as random events, giving a more formal meaning to the notion of ``quantum randomness''.
However, given the intricacies of randomness, it should be clear that this refers to the measurement \emph{process}, and does not entail that quantum measurement outcomes are maximally random.
In fact, maximal randomness in the sense that no correlations exist between successive measurement results is mathematically impossible~\cite{GS-90,calude:02}: there exist only degrees of randomness with no upper limit.
As a result, any claims regarding the quality of quantum randomness need to be analysed carefully.

Indeed, in many applications of quantum randomness stronger computability theoretic notions of randomness, such as Martin-L\"of randomness~\cite{calude:02}, which apply to sequences of outcomes would be desirable.
\emph{It is not known if quantum outcomes are indeed random in this respect.}
However, it was shown previously \cite{svozil-2006-ran,2012-incomput-proofsCJ} that \emph{a sequence $\x$ produced by repeated outcomes of a value indefinite observable must be bi-immune.}\footnote{See Sect.~\ref{sec:alg} for definitions.}
This result was proved using a further physical assumption, related to and motivated by the EPR principle, called the \emph{e.p.r. assumption}.\footnote{Here, e.p.r. stands for `elements of physical reality, not `Einstein, Podolsky and Rosen' as in the EPR principle.'} 
This assumption states that, \emph{if a repetition of measurements of an observable generates a computable sequence, then this implies these observables were value definite prior to measurement.}
In other words, it specifies a particular sufficient condition for value definiteness.

Given the relation between unpredictability and Tadaki total unpredictability (which implies bi-immunity) discussed in Sect.~\ref{sec:alg}, it is natural to ask whether the bi-immunity of sequences generated by measuring repeatedly a value indefinite observable 
 is a general consequence of its unpredictability, or if it is an independent consequence of value indefiniteness.

The links between unpredictability and Tadaki total unpredictability we explored earlier are relative to the use of specific extractors---such as $\xi_p$---and, as we discussed, need not hold when other more physically relevant extractors are considered.
Furthermore, for the unpredictability of an experiment $E$ to guarantee that \emph{any} outcome of an infinite repetition of $E$ be incomputable---a much weaker statement than bi-immunity---it would have to be the case that (taking the contrapositive) if even a single infinite repetition $\lambda_1,\lambda_2,\dots$ of $E$ could generate a computable sequence this would imply that $E$ is predictable.
However, the definition of a predictor $P_E$ for $E$ requires that $P_E$  gives correct predictions for \emph{all} repetitions.
Hence, we will elaborate a simple example of an unpredictable experiment $E$ that can produce \emph{both} computable and incomputable sequences, showing that unpredictability does not imply incomputability (let alone bi-immunity).

Let $d$ be the dyadic map; that is, the operation on  infinite sequences of bits defined by $d(x_1 x_2 x_3\dots)= x_2 x_3\dots$.
This operation is well known to be chaotic and equivalent (more precisely, topologically conjugate) to many others, e.g.~the logistic map with $r=4$~\cite{Devaney-1989}.
Let us consider an experiment $E_d$ which involves iterating the dyadic map $k\ge 2$ times on a `seed' $\x=0 x_2 x_3\dots$ until  $x_{k+1}=0$. 
In other words, given $\x$ we look for 
the smallest integer $k\ge 2$ such that $x_{k+1}=0$, hence $d^k(\x)=0 x_{k+2} x_{k+3}\dots$.
\emph{If such a $k$ exists, then the outcome of the experiment is}  $x_{k+2}\in\{0,1\}$.
\emph{We assume that such an $E_d$ (ideally) is physically implementable.}
We have chosen this example for simplicity;
a more `physically natural' example might be the evolution of a chaotic double pendulum from some set initial condition (up to finite accuracy) for which the outcome is read off once the pendulum returns sufficiently close to its initial conditions.

This experiment can, of course, be repeated in many different ways to generate  an infinite sequence, but it suffices to consider the simplest case where the transformed seed $\x^{(1)}=d^k(\x)$ after one iteration is taken as the seed for the next step; note that this, by design, satisfies the requirement that the first bit of $\x^{(1)}$ is 0 (i.e., $x_1^{(1)}=0$), provided $k$ exists.
\emph{Let us assume further that any sequence $\x=x_1x_2\dots$ such that $x_1=0$ is a valid physical seed}. For the case of a double pendulum this is akin to assuming that the position of a pendulum can take any value in the continuum---not an unreasonable, if nonetheless  important, assumption.

Let $\y=y_1y_2\dots$ be an arbitrary infinite sequence, and consider the sequence $\x=010y_10y_20y_3\dots$.
For any such sequence $\x$ of this form, $d^2(\x)=0y_10y_2\dots$, so the outcome of $E_d$ with seed $\x$ is precisely $y_1$, and the new seed $\x^{(1)}=d^2(\x)=0y_10y_2\dots$.
Similarly, for all $i$, starting with the seed $\x^{(0)}=\x$, the outcome of the $i$th repetition is precisely $y_i$, since a minimum number of $k=2$ applications of $d$ suffices for the first bit of $d^2(\x^{(i-1)})$ to be 0, and the seed after this repetition is precisely $\x^{(i)}=0y_i0y_{i+1}\dots$.
Hence, starting with the seed $\x$ one obtains the infinite sequence $\y$ by repeating $E_d$ to infinity.
In particular, since $\y$ can be any sequence at all, one can obtain both computable and incomputable sequences by repeating $E_d$.

Let us show also that $E_d$ is unpredictable.
Let us assume, for the sake of contradiction, that there exists a predictor $P_{E_d}$ and extractor $\xi_d$ such that $P_{E_d}$ is correct for $\xi_d$.
Then $P_{E_d}$ must give infinitely many correct predictions using $\xi_d$ for any two runs $\lambda_1\lambda_2\dots$ and $\lambda_1'\lambda_2'\dots$ which differ only in their seeds $\x$ and $\x'$.
In particular, this is true if $\x,\x'$ are sequences of the form $0a_1a_2\dots$ where $a_i\in\{1^t00,1^t01\}$ for all $i$, and $t\ge 1$ is fixed, since these are possible seeds for $E_d$.
For such seeds $\x,\x'$ the minimum $k\ge 2$ such that the first bit of $d^k(\x)$ is 0 is precisely $k=t+1$.
Furthermore, if we let $\x^{(0)}=\x$ and $\x^{(i)}=d^{k_i}\left(\x^{(i-1)}\right)$ be the seed for the $i$th repetition of $E_d$, then $k_i=t+1$ for all $i$; that is, each iteration of $E_d$ shifts the seed precisely $t+1$ bits.
Thus, to make infinitely many predictions for $E_d$ starting with seeds $\x$ and $\x'$ correctly, $P_E$ must have access, via $\xi_d$, to more than $t+3$ bits of the current seed, since the first $t+2$ bits of $\x^{(i)}$ and $\x'^{(i)}$ are the same for all $i$.
However, since $t$ is arbitrary, and the same extractor $\xi_d$ must be used for all repetitions regardless of the seed, this implies that $\xi_d$ is \emph{arbitrarily accurate}, which it is, again, not unreasonable  \emph{to assume to be physically impossible}. Consequently, $E_d$ must be unpredictable.

The construction of $E_d$  may be slightly artificial and its unpredictability relies, of course, on certain physical assumptions about the possibility of certain extractors.
However, this concrete example shows that there is no mathematical obstacle to an unpredictable experiment producing both computable and incomputable outcomes when repeated, and is, at the very least, physically conceivable.

Any link between the unpredictability of an experiment and computability theoretic properties of its output thus relies critically on physical properties---and assumptions---of the particular experiment.
Indeed, this careful dependance on the particular physical description of $E$ is one of the strengths of this general model. This gives the model more physical relevance as a notion of (un)predictability than purely algorithmic proposals.

The bi-immunity of quantum randomness is a crucial illustration of this fact.
Using a slightly a stronger additional 
hypothesis on the nature of value (in)definiteness,  bi-immunity can be guaranteed for every sequence of quantum random bits obtained by measuring a value indefinite observable~\cite{2012-incomput-proofsCJ}.
 For this particular quantum experiment bi-immunity complements, and is  independent of, unpredictability.\footnote{Recall that   bi-immunity need not imply unpredictability either.}

\section{Summary}

In this paper, we addressed two specific points relating to physical unpredictability.
Firstly, we developed a generalised model of prediction for both individual physical events, and (by extension) infinite repetitions thereof.
This model formalises the notion of an effective prediction agent being able to predict `in principle' the outcome of an effectively specified physical experiment.
This model can be applied to classical or quantum systems of any kind to assess their (un)predictability, and doing so to various systems, particularly classical, could be an interesting direction of research for the future.

Secondly, we  applied this model to quantum measurement events.
Our goal was to formally deduce the unpredictability of single quantum measurement events, via the strong Kochen-Specker theorem and value indefiniteness, rather than rely on the \emph{ad hoc} postulation of these properties.

More specifically,  suppose that we prepare a quantum in a pure state corresponding to a unit vector in Hilbert space of dimension at least three.
Then any complementary observable property of this quantum---corresponding to some projector whose respective linear subspace is neither collinear nor orthogonal with respect to the pure state vector---is value indefinite.
Furthermore,  the outcome of a measurement of such a property is unpredictable with respect to our model of prediction.

Quantum value indefiniteness is key for the proof of unpredictability.
In this framework, the bit resulting from the measurement of such an observable property is ``created from nowhere'' (\emph{creatio ex nihilo}), and cannot be causally connected to any physical entity, whether it be knowable in practice or hidden.
While quantum indeterminism is often informally treated as an assumption in and of itself, it is better seen as a formal consequence of Kochen-Specker theorems in the form of value indefiniteness.
(Indeed, without these theorems such an assumption would appear weakly grounded.)
Yet this derivation of value indefiniteness rests on the three assumptions: admissibility, non-contextuality, and the eigenstate principle.
As we discussed in Sect.~\ref{sec:ac}, models in which some of these assumptions are not satisfied exist.

The single-bit unpredictability of the output obtained by measuring a value indefinite quantum observable 
complements the fact---proven in~\cite{2012-incomput-proofsCJ} with an additional hypothesis---that such an experiment generates, in the limit, a strongly incomputable sequence.
We show that this additional hypothesis is necessary in the sense that unpredictable experiments are, in general, capable of generating both incomputable and computable infinite sequences.

The unpredictability and strong incomputability of these quantum measurements ``certify'' the use of the corresponding quantum random number generator for various computational tasks in cryptography and elsewhere~\cite{svozil-qct,stefanov-2000,10.1038/nature09008}. As a consequence, this quantum random number generator can be seen and used as an \emph{incomputable oracle}, thus  justifying   a form of  \emph{hypercomputation}. Indeed,  no universal Turing machine can ever produce in the limit an output that is identical with the sequence of bits generated by this quantum oracle~\cite{qrand-oracle}.
More than that---no single bit of such sequences can ever be predicted. 
Evaluating the computational power of a (universal) Turing machine provided with a quantum random oracle certified by maximum unpredictability is a challenging, both theoretical and practical, {\it open problem}.

In this context incomputability appears \emph{maximally} in two forms: \emph{individualised}---no single bit can be predicted with certainty (Theorem~\ref{unpredictsingle}); that is,  an algorithmic computation
of a single bit, even if correct, cannot be formally certified; and, relative to slightly stronger hypotheses, \emph{asymptotic} via  bi-immunity---only finitely many bits can be correctly predicted via an algorithmic computation.

Finally, we  emphasise that the indeterminism and unpredictability of quantum measurement outcomes proved in this paper are based on the  strong form of the Kochen-Specker theorem, and hence require at minimum three-dimensional Hilbert space.
The question of whether this result can also be proven for two-dimensional Hilbert space without simply assuming value indefiniteness is an \emph{open problem};
this question is important not only theoretically, but also practically, because many current quantum random generators are based on two-dimensional  measurements.

\section*{Acknowledgement} 
Abbott thanks Thierry Paul for discussions on the assumptions needed to deduce quantum incomputability.
Calude thanks Giuseppe Longo for raising the question of the randomness of individual bits
(September 2008).
This work was supported in part by Marie Curie FP7-PEOPLE-2010-IRSES Grant RANPHYS.

\bibliographystyle{splncs03}

\end{document}